\documentclass[onecolumn]{elsart}
\usepackage{amssymb}
\usepackage{graphicx}
\usepackage{enumerate}
\begin{document}

\begin{frontmatter}
\title{Reply to: `Reply to: ``Comment on: `How much security does Y-00 protocol provide us?''''}
\author{Ranjith Nair}\ead{r-nair@northwestern.edu},
\author{Horace P. Yuen},
\author{Eric Corndorf},
\author{Prem Kumar}
\address{Center for Photonic Communication and Computing\\
Department of Electrical and Computer Engineering\\
Department of Physics and Astronomy\\
Northwestern University, Evanston, IL 60208}

\begin{abstract}
Nishioka \emph{et al} claim in \cite{nishioka05}, elaborating on
their earlier paper \cite{nishioka04}, that the direct encryption
scheme called Y-00 \cite{prl,yuen04} is equivalent to a classical
non-random additive stream cipher, and thus offers no more
security than the latter. In this paper, we show that this claim
is false and that Y-00 may be considered equivalent to a
\emph{random} cipher. We explain why a random cipher provides
additional security compared to its nonrandom counterpart. Some
criticisms in \cite{nishioka05} on the use of Y-00 for key
generation are also briefly responded to.
\end{abstract}
\end{frontmatter}

\section{Introduction}

The direct encryption system called Y-00 or $\alpha\eta$
\cite{prl,yuen04} uses coherent states to transmit encrypted
information between two users, Alice and Bob, sharing a secret
key. Nishioka \emph{et al} claimed in \cite{nishioka04} that the
security of Y-00 was completely equivalent to that of a classical
non-random additive stream cipher. Although we rebutted this claim
in \cite{pla05}, the authors of \cite{nishioka04} have replied in
\cite{nishioka05} to the effect that we did not understand the
claims made in their original paper, and that our reply was
irrelevant. It is in fact true that some details of our
understanding of the claims in \cite{nishioka04} differ from the
purported clarifications of the same provided in
\cite{nishioka05}. However, setting aside questions of clarity of
the claims in \cite{nishioka04}, we understand their claims in
\cite{nishioka05} exactly and contend that the claimed equivalence
of Y-00 to a classical non-random additive stream cipher is false,
thus reiterating the conclusion of \cite{pla05}. We provide
arguments in this paper that conclusively demonstrate that Y-00 is
not equivalent to a non-random cipher.

In Section 2, we briefly review the Y-00 direct encryption
protocol. In Section 3, we describe the claims in
\cite{nishioka04} and \cite{nishioka05}. In Section 4, we review
the concept of a random cipher and describe why they are more
secure than non-random ones against attacks on the key, thus
highlighting the added feature that Y-00 provides over a
non-random cipher. We then rebut the claims in \cite{nishioka05}
and show why Y-00 is not equivalent to a non-random stream cipher.
In Section 5, we briefly respond to the criticisms in
\cite{nishioka05} of using Y-00 with weak coherent states as a key
generation system.

\section{The Y-00 Direct Encryption Protocol}

We review the Y-00 protocol, using the notations of
\cite{nishioka05} for easy reference.
\begin{enumerate}[(1)]

\item
Alice and Bob share a secret key $\mathbf{K}_s$.

\item
Using a pseudo-random-number generator $PRNG(\centerdot)$, e.g., a
linear feedback shift register, the seed key $\mathbf{K}_s$ is
expanded into a running key sequence
$\mathbf{K}=PRNG(\mathbf{K}_s)=(K_1, \ldots , K_N)$, with each
block $K_i \in \{0, 1, \ldots, M-1\}$.

\item
For each bit $r_i$ of a plaintext sequence $\mathbf{R}_N = (r_1,
\ldots, r_N)$, Alice transmits the coherent state
\begin{equation} \label{state}
|\psi(K_i, r_i)\rangle=|\alpha e^{i\theta(K_i,r_i)}\rangle,
\end{equation}
where $\alpha \in \mathbb{R}$ and
$\theta(K_i,r_i)=[K_i/M+(r_i\oplus\Pi(K_i))]\pi$. $\Pi(K_i)= 0$ or
$1$ according to whether $K_i$ is even or odd. This distribution
of possible states is shown in Fig. 1. Thus $K_i$ can be thought
of as choosing a `basis' with the states representing bits $0$ and
$1$ as its end points.

\item
Bob, knowing $K_i$, makes a measurement to discriminate just the
two states $|\psi(K_i, r_i)\rangle$ and $|\psi(K_i, r_i \oplus
1)\rangle$.
\end{enumerate}

\begin{figure*}\begin{center}
\includegraphics[width=13cm]{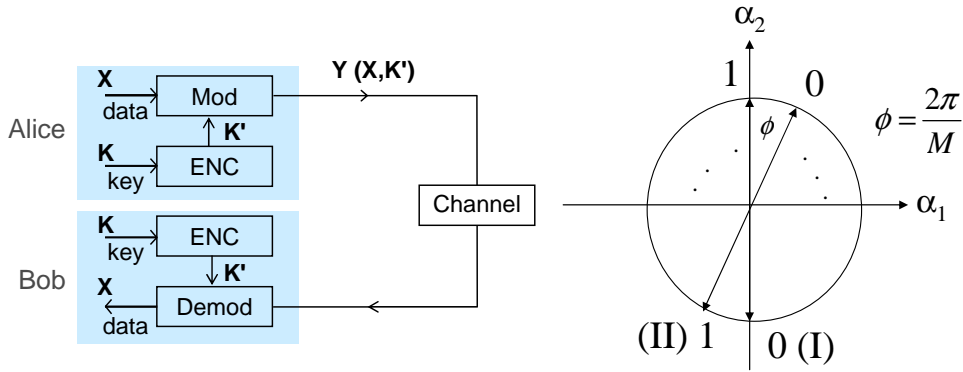}
\caption{Left: Overall schematic of the $\alpha\eta$ scheme.
Right: Depiction of $M/2$ bases with interleaved logical state
mappings.}\label{setup}
\end{center}\end{figure*}

The probability that Bob makes an error can be made negligibly
small by choosing the mean photon number $S\equiv |\alpha|^2$
large enough. In particular, the optimal quantum measurement
\cite{helstrom76} for Bob has error probability
\begin{equation} \label{pB}
P^B_e \sim \frac{1}{4} exp(-4S).
\end{equation}

Eve, not knowing $K_i$, is faced with the problem of
distinguishing the density operators $\rho^0$ and $\rho^1$ where
\begin{equation} \label{rho}
\rho^b=\sum_{K_i}\frac{1}{M}|\psi(K_i,b)\rangle\langle\psi(K_i,b)|.
\end{equation}
For a fixed signal energy $S$, Eve's optimal error probability is
numerically seen to go asymptotically to $1/2$ as the number of
bases $M \rightarrow \infty$ (See Fig. 1 of \cite{prl}). The
intuitive reason for this is that increasing $M$ more closely
interleaves the states on the circle representing bit 0 and bit 1,
making them less distinguishable. Therefore, at least under
\emph{individual} attacks on each qumode, the Y-00 protocol offers
any desired level of security determined by the relative values of
$S$ and $M$.

One can also ask if Eve can obtain the key $\mathbf{K}_s$ under a
\emph{known-plaintext} attack, thus compromising the security of
future data. While the complete analysis of the security of Y-00
under known-plaintext analysis has not been performed, we can
still make some remarks about the security that Y-00 offers
against an attack involving a fixed measurement (e.g., a
heterodyne or phase measurement) on each qumode followed by a
\emph{brute-force trial} of remaining key candidates. Indeed, a
simple estimate of the noise in the phase measurement (which
performs better than the heterodyne measurement) can be obtained
by assuming that the noise moves the measured angle around the
transmitted value uniformly within a standard deviation
$1/|\alpha|$ of the measurement noise. Then, it is seen that the
number of possible bases $N_\sigma$ consistent with the known bit
$b$ in each measurement is $N_\sigma = M/(2\pi|\alpha|)$. Thus,
the randomization provided by the quantum noise creates a search
problem for Eve that would \emph{not} be present if Y-00 was a
non-random cipher.

\section{Claims in Nishioka \emph{et al} \cite{nishioka05}}

Nishioka \emph{et al} claim that Y-00 can be reduced to a
classical non-random stream cipher under the attack that we now
review. For each transmission $i$, Eve makes a heterodyne
measurement on the state and collapses the outcomes to one of $2M$
possible values. Thus, the outcome $j \in \{0, \cdots, 2M-1\}$ is
obtained if the heterodyne result falls in the wedge for which the
phase $\theta \in [\theta_j-\pi/2M, \theta_j+\pi/2M]$, where
$\theta_j= \pi j/M$. Further, for $q \in \{0, \cdots, M-1\}$
representing the $M$ possible values of each $K_i$, Nishioka
\emph{et al} construct a function $F_j(q)$ with the property that,
for each $i$, and the corresponding running key value $K_i$
actually used,
\begin{equation} \label{decryption}
F_{j^{(i)}}(K_i)=r_i
\end{equation}
with probability very close to 1. In fact, for the parameters
$S=100$ and $M=200$, they calculate the probability that
Eq.(\ref{decryption}) fails to hold to be $10^{-44}$, which value
they demonstrate to be negligible for any practical purpose.

The authors of \cite{nishioka05} further claim that the above
function $F_{j^{(i)}}(q)$ can always be represented as the XOR of
two bit functions $G_{j^{(i)}}(q)$ and $l_{j^{(i)}}$, where
$l_{j^{(i)}}$ depends \emph{only} on the measurement result. Thus,
they make the claim that the equation
\begin{equation} \label{reduction}
l_{j^{(i)}}=r_i \oplus G_{j^{(i)}}(K_i)
\end{equation}
holds with probability effectively equal to 1. They then observe
that a classical additive stream cipher \cite{mvv96,massey88}
(which is non-random by definition) satisfies
\begin{equation}\label{streamcipher}
l_i=r_i \oplus \tilde{k_i},
\end{equation}
where $r_i$, $l_i$, and $\tilde{k_i}$ are respectively the $i$th
plaintext bit, ciphertext bit and running key bit. Here,
$\tilde{k_i}$ is obtained by using a seed key in a
pseudo-random-number generator to generate a longer running key.
The authors of \cite{nishioka05} then argue that since
$l_{j^{(i)}}$ in Eq.(\ref{reduction}), like the $l_i$ in
Eq.(\ref{streamcipher}), depends just on the measurement result,
the validity of Eq.(\ref{reduction}) proves that the security of
Y-00 is equivalent to that of a classical stream cipher. In
particular, they claim that by interpreting $l_{j^{(i)}}$ as the
ciphertext, Y-00 is not a random cipher, i.e., it does not satisfy
Eq.(\ref{random}) of the next section.

We analyze and respond to these claims and other statements in
\cite{nishioka05} in the following section.

\section{Reply to claims in \cite{nishioka05}}

First of all, we review the definition of a \emph{random} cipher.
Such a cipher is called a `privately-randomized cipher' in
\cite{massey88}, but we will call it just a random cipher. A
random cipher is defined by the two conditions:
\begin{equation} \label{random}
H(Y_N|\mathbf{K}_s, R_N)\neq 0,
\end{equation}
and
\begin{equation} \label{decrypt}
H(R_N|Y_N,\mathbf{K}_s)=0.
\end{equation}
Here, $Y_N$ refers to the $N-$ symbol long ciphertext and $R_N$
and $\mathbf{K}_s$ are the plaintext and secret key, as in section
1. Note that there is no restriction on the alphabet of $Y_N$,
which may be binary, $M$-ary or even continuous. Eq.(\ref{random})
implies that, for a given key, the plaintext may be mapped by the
encrypter into more than one possible ciphertext. However, it is
still required that the plaintext can be recovered from the
ciphertext and the key, which is the meaning of
Eq.(\ref{decrypt}). The case where Eq.(\ref{decrypt}) holds but
Eq.(\ref{random}) does not is the usual case of a non-random
cipher in standard cryptography.

The advantage of a random cipher, which is briefly described in \cite{pla05} but not appreciated in \cite{nishioka05}, is that it may be secure against
attacks on the key in the case when the attacker knows the
statistics $p(R_N)$ of the data. In the case where the $r_i$ are
independent and identically distributed, a random cipher can be
constructed that provides complete information-theoretic security
of the key \cite{gunther88}, in the sense that
$H(\mathbf{K}_s|Y_N)=H(\mathbf{K}_s)$. Such security cannot be
obtained with nonrandom ciphers \cite{yuen05}.  See \cite{yuen05}
for a detailed discussion on the security of random ciphers.
Although we do not claim such information-theoretic security for
Y-00, the possibility is not ruled out. We have already commented
on the added brute-force search complexity of Y-00 against attacks
on the key in Section 2. We now proceed to prove that the claim in
\cite{nishioka05} that Y-00 is reducible for Eve to a non-random
stream cipher under a heterodyne measurement is false.

To begin with, we believe that Eq.~(\ref{decryption}) (Eq.~(14) in
\cite{nishioka05}) is correct with the probability given by them.
The content of this equation is simply that Eve is able to
decrypt the transmitted bit from her measurement data $J_N$ and
the key $\mathbf{K}_s$. In other words, it merely asserts that
Eq.(\ref{decrypt}) holds for $Y_N=J_N$. As such, it does not
contradict, and is even \emph{necessary}, for the claim that Y-00
is a random cipher for Eve. In fact, we already claimed in
\cite{yuen04} and \cite{pla05} that such a condition holds. In
this regard, note also that the statement in Section 4.1 of
\cite{nishioka05} that ``informational secure key generation is
impossible when ( Eq.(\ref{decryption}) of this paper) holds'' is
irrelevant, since direct encryption rather than key generation is
being considered here.

We also agree with the claim of Nishioka \emph{et al} that it is
possible to find functions $l_{j^{(i)}}$ and $G_{j^{(i)}}(q)$, the
former depending only of the measurement result $j^{(i)}$, such
that Eq.(\ref{reduction}) holds, again with probability
effectively equal to one. The error in \cite{nishioka05} is to use
this equation to claim, in analogy with Eq.~(\ref{streamcipher}),
that Y-00 is reducible to a classical stream cipher, and hence
non-random.

To understand the error in their argument, note that, for
Eq.~(\ref{streamcipher}) to represent an \emph{additive} stream cipher,
the $l_i$ in that equation should be a function
\emph{only} of the measurement result, and $\tilde{k_i}$ should be
a function \emph{only} of the running key. While the former
requirement is true also for the $l_{j^{(i)}}$ in
Eq.~(\ref{reduction}), the latter is certainly \emph{false} for
the function $G_{j^{(i)}}(K_i)$ in Eq.~(\ref{reduction}), since it
depends \emph{both} on the measurement result $j^{(i)}$ and the
running key $K_i$. Indeed, it can be seen that the definition of the function
$F_{j^{(i)}}(K_i)$, and thus, $G_{j^{(i)}}(q)$ depends on the sets
$C_{j^{(i)}}^+$ and $C_{j^{(i)}}^+$ defined in Eq.~(12) of
\cite{nishioka05}. The identity of these sets in turn depends on
the relative angle between the basis $q$ and Eve's estimated basis
$\tilde{j^{(i)}}= j^{(i)} \bmod M.$ Thus, it is clearly the case
that $G_{j^{(i)}}(K_i)$ must depend both on $j^{(i)}$ and $K_i$, a
fact also revealed by the inclusion of the subscript $j^{(i)}$ by
the authors of \cite{nishioka05} in the notation for $G$.

We have shown above why the representation of Y-00  via Eq.~(\ref{reduction})
is not equivalent to an additive stream cipher. The question may be raised,
however, if  Eq.~(\ref{reduction}) reduces Y-00 to any kind of nonrandom cipher whatsoever.
We will show below that the answer is negative. Indeed, Nishioka \emph{et al} emphasize that
Y-00 is nonrandom because
\begin{equation} \label{l}
H(L_N|R_N, \mathbf{K}_s) =0
\end{equation}
holds, where $\mathbf{L}_N=(l_{j^{(1)}}, \ldots, l_{j^{(N)}})$.
This equation follows from Eq.~(\ref{reduction}) and so by
considering $\mathbf{L}_N\equiv Y_N$ to be the ciphertext, the
Eq.(\ref{random}) is not satisfied, thus supposedly making Y-00
nonrandom. 
The choice of $\mathbf{L}_N$ as the ciphertext is supported by the statement in
\cite{nishioka05} that ``It is a matter of preference what we should refer to
as ``ciphertext''.''  This is not true without qualification. It ignores the \emph{crucial} point that the random variable 
that is chosen
as ciphertext must be sufficient to decrypt to the corresponding $R_N$ for every value of
the key. We will show below that, for Y-00, the ciphertext alphabet needs to be atleast
$(2M)$-ary, although larger, even continuous alphabets (such as the possible values of phase angle in a phase measurement or
the two quadrature amplitudes in a heterodyne measurement) may be used. Thus, if one wants
to claim equivalence to a classical cipher (random or non-random), for a particular choice of ciphertext $Y_N$,
one must show that  Eq.~(\ref{decrypt}) is
satisfied using that \emph{same} ciphertext $Y_N$. In the case where $Y_N=\mathbf{L}_N$, one
may readily see that  Eq.~(\ref{decrypt}) is not satisfied, i.e.,
 $H(\mathbf{R}_N|\mathbf{L}_N,\mathbf{K}_s)\neq 0$.
The reason is that, as we noted in our description above of the
function $G_{j^{(i)}}(q)$, decrypting $r_i$ requires knowledge of
certain ranges in which the angle between the basis chosen by the
running key and the estimated basis $\tilde{j^{(i)}}$ falls. To
convey this information \emph{for every possible} $j^{(i)}$, one
needs at least $\log_2(2M)$ bits. It follows that the single bit
$l_{j^{(i)}}$ is insufficient for the purpose of decryption, and
so Eq.~(\ref{decrypt}) cannot be satisfied for $Y_N=\mathbf{L}_N$.
Therefore, we conclude, that in the interpretation of
$\mathbf{L}_N$ as the ciphertext, decryption is not possible even
if Eve has the key $\mathbf{K}_s$. Indeed, it is $\mathbf{J}_N$
that can be regarded as a possible ciphertext, since
Eq.~(\ref{decrypt}) is satisfied for $Y_N=\mathbf{J}_N$. The fact that $\mathbf{J}_N$
is a true ciphertext sufficient for decryption is implicit in the dependence on $j^{(i)}$ of the function $G$
in Eq.~(\ref{reduction}).  However,
with this choice of ciphertext, Y-00 necessarily becomes a
\emph{random} cipher, because
$H(\mathbf{J}_N|\mathbf{R}_N,\mathbf{K}_s) \neq 0$ -- this latter fact
is
admitted by Nishioka \emph{et al} in \cite{nishioka05}.

We hope that the discussion above makes it clear that the
`reduction' of Y-00 in \cite{nishioka05} to a non-random cipher is
false, and that in fact, no such reduction can be made under the
heterodyne attack considered in \cite{nishioka05}. However, the
representation of ciphertext by $Y_N=\mathbf{J}_N$ does reduce it
to a \emph{random} cipher under the heterodyne attack.
As a result, it can be implemented classically in principle, but
not in practice. This is because true random numbers can
only be generated physically, not by an algorithm, and the
practical rate for such generation is many orders of magnitude
below the $\sim$ Gbps rate in our experiments where the
coherent-state quantum noise does the randomization
\textit{automatically}. Furthermore, our physical ``analog''
scheme does not sacrifice bandwidth or data rate compared to other
known randomization techniques.
See \cite{yuen05} for a detailed discussion.

We conclude this section by responding to some other statements
made in \cite{nishioka05}.

In Section 3,2, Nishioka \emph{et al} state that ``It is interesting to note
that a smaller $M$ (but not $M$=1) is preferable for increasing the stochastic
property.'' Here, they mean that the decryption using $\mathbf{J}_N$ and the
key is noisier for smaller $M$. We claim that this cannot be the case and
that the decryption probability is essentially independent of $M$. In any case,
for the heterodyne quantum noise to cover more states on the circle, it is clear
that a larger $M$ is preferable (See our discussion in Section 2). 

In Section 3.3, Nishioka \emph{et al} claim that ``The value of
$l_{j^{(i)}}$ does not have to be the same as that of
$l_{j^{(i')}}$ when $i \neq i'$, even if $j^{(i)}=j^{(i')}$
holds.'' This statement is in direct contradiction to their
previous statement in the same subsection that ``$l_{j^{(i)}}$
depends only on the measurement value $j^{(i)}$''.

In the same subsection, Nishioka \emph{et al} claim that ``In
(\cite{nishioka04}), we showed another concrete construction of
$l_{j^{(i)}}$ ...''. In our opinion, there was no explicit
construction of $l_{j^{(i)}}$ in that paper, which to us seemed
quite vague. We were led to the choice of $l_i$ described in
\cite{pla05} by the attempt to make the additive stream cipher
representation Eq.~(\ref{streamcipher}) valid. In fact, such a
representation is claimed by Nishioka \emph{et al} in their Case 2
of \cite{nishioka04}. It turned out, however, that decryption
using that $l_i$ suffered a $0.1-1$\% error depending on the value
of $S$ used. See \cite{pla05} for further details. In any case, as
we have shown above, no construction of a single-bit from the
heterodyne measurement results can satisfy Eq.(\ref{decrypt}) with
the extremely low probability given in \cite{nishioka05}.

\section{Remarks regarding Key Generation using Y-00}

In \cite{pla05}, we replied to the claim that
information-theoretically secure key generation is impossible for
Y-00 by showing a 6 dB advantage that the users have over Eve
launching a heterodyne attack. This advantage can be used for
practical key generation using a small enough value of $S$. This
is acknowledged in \cite{nishioka05}, thus validating our claim
that it is indeed possible. However, the new issue is raised in
\cite{nishioka05} that this advantage is too small to allow Y-00
to generate keys, in their example, over a distance of 50 km. In
this connection, we merely wish to state that, (i) this is not the
original issue under dispute and we do not wish to bring a new
issue into the present discussion; (ii) similar loss limits are
also present for BB84; (iii) other techniques and schemes are
already discussed in \cite{yuen04} to overcome this distance
limit.

\section{Conclusion}

We have demonstrated that, under a heterodyne measurement, the
Y-00 Direct Encryption protocol cannot be reduced to a classical
non-random stream cipher, as claimed in
\cite{nishioka05}. Its security under heterodyne attack is
equivalent to a corresponding random cipher.

\bibliographystyle{elsart-num}

\end{document}